# Growth by Intervention: State Demand and Preferential Allocation in the Hungarian Construction Sector since 2010

Zoltán Bartha https://orcid.org/0000-0002-9208-7881

Abstract

This paper examines the post-2010 expansion of the Hungarian construction sector as a case study of large-scale government intervention in a single market. A political economy framework conceptualises the intervention as two mutually reinforcing policies — expansionary demand (public procurement and housing subsidies) and preferential allocation towards politically connected firms — operating under inelastic supply through two mechanisms: market-level subsidy capitalisation and firm-level rent extraction. Using a structured descriptive case study design, the paper draws on Hungarian Central Statistical Office and Eurostat macroeconomic statistics, firm-level data on Hungary's largest companies (CrefoPort), policy documents, and a regional comparison to assess four propositions derived from the framework. The evidence supports all four: real construction output grew faster than in any regional peer except Romania; construction producer prices persistently outpaced the GDP deflator; ownership of the leading firms shifted towards politically connected domestic owners, whose profitability rose sharply while their exports remained negligible; labour productivity per hour worked returned by 2024 to roughly its 2000 level; and connected firms captured a rising share of sector profits, as the subsidies' value was partly eroded by the price increases they induced. Competing explanations do not jointly account for this configuration. The findings illustrate how combined demand stimulus and favouritism produce extensive growth without efficiency gains.

## 1. Introduction

Hungary's construction sector expanded rapidly from the second half of the 2010s. While the post-2008 period had left the sector badly weakened, the subsequent recovery made construction one of the main beneficiaries of the upturn. Its share of gross value added had fallen to 3.9% by 2012; by 2022 it had risen to 6.4%, the highest level since 1990 (Hungarian Central Statistical Office — KSH). Employment followed the same pattern, dropping below 247,000 in 2012 before rebounding by over 50% to around 390,000 by 2023–24 (KSH). By any conventional measure of scale — output, employment, share of the economy — the sector's trajectory was a success story.

A closer look reveals a more puzzling picture. The expansion was accompanied by exceptional price dynamics: according to Eurostat's House Price Index, Hungarian residential property prices rose faster than anywhere else in the EU after 2015, and construction producer prices persistently outpaced the general price level. Meanwhile, labour productivity in the sector did not improve (Eurostat). Profitability, by contrast, rose above the EU average (European Commission, 2023, p. 56), even as the leading firms' export activity remained negligible. This configuration motivates the central question of the paper: how can a sector grow this rapidly while its productivity stagnates, its prices outpace the rest of the economy, and its profitability becomes decoupled from any conventional indicator of competitiveness — and what role did state intervention play in producing this outcome?

The Hungarian construction sector is examined here as a case study of a more general theoretical problem: the economic consequences of unusually large-scale government intervention in a single market. The problem is not country-specific. Demand-side housing programmes, infrastructure stimulus, and cohesion-funded public investment are standard instruments across developed and emerging economies, and political favouritism in public procurement is a well-documented global phenomenon. What makes the Hungarian case particularly valuable is that both forms of intervention occurred simultaneously, at exceptional scale, over an extended period, and in a single, observable sector: since 2015, at least HUF 4,500 billion was channelled into household home creation through subsidies and subsidised loans (Makay et al., 2024), while public investment between 2010 and 2025 — much of it construction — exceeded HUF 30,000 billion according to government statements (Profitline, 2025). Few settings offer a comparably clear view of what happens when massive demand stimulus meets politically structured supply.

The paper develops a simple political economy framework built on two interconnected interventions — expansionary demand policy and the preferential allocation of that demand to politically connected firms — whose interaction, under inelastic supply, is argued to produce the observed pattern. The framework and its propositions are set out in Section 3. The main contribution of the paper is that it connects the housing economics literature on subsidy capitalisation, the empirical literature on politically connected firms, and the political economy literature on post-2010 Hungary. These three strands have largely developed in isolation, but the model suggested in this paper merges them into a single sectoral account in which overheating, concentrated profitability, and stagnant productivity emerge as joint consequences of the same policy mix.

Methodologically, the paper is a structured descriptive case study. It draws on macroeconomic and sectoral statistics from the KSH and Eurostat, firm-level data on Hungary's largest companies from the CrefoPort database, official policy documents and central bank reports, and a comparison with peer economies from the region. The analysis assesses whether the observed evidence is consistent with the framework's predictions and whether competing explanations account for the same patterns equally well; it does not estimate causal effects econometrically, and its claims are bounded accordingly.

The evidence is consistent with all four propositions. Hungary's real construction output growth places it among the region's strongest expanders, exceeded only by Romania and far above the other Visegrád economies; sectoral prices diverged from the general price level early and persistently, in line with the capitalisation mechanism; ownership of the leading firms shifted decisively towards politically connected domestic owners, whose profitability rose sharply while their exports remained negligible; and productivity stagnated throughout — a combination that competing explanations do not jointly account for.

The remainder of the paper is organised as follows. Section 2 describes the data and methodology. Section 3 sets out the theoretical framework, its propositions, its relation to the existing literature, and the competing explanations considered. Section 4 presents the macroeconomic evidence, including the regional comparison. Section 5 examines the three main policy instruments — housing subsidies, the preferential VAT rate, and public investment. Section 6 documents the transformation of the sector's ownership structure. Section 7 analyses the evolution of productivity and the character of connected firms' expansion. Section 8 assesses the evidence on rent-seeking. Section 9 concludes.

## 2. Data and Methodology

This study employs a structured descriptive case study approach. The analytical framework draws on political economy theory, treating sectoral outcomes — prices, productivity, profitability, and ownership concentration — as the joint product of market forces and policy choices, including interventions that selectively shape market access in favour of politically connected actors. This framing draws on established literature on subsidy capitalisation (Hilber, 2017), public procurement and favouritism (Fazekas et al., 2015), and rent-seeking in post-communist economies (Celis Galvez és mtsai., 2025; Szanyi, 2022). A single-country, single-sector design is appropriate because the mechanisms under investigation — specific policy instruments, procurement practices, and ownership transformations — are inherently country-specific and cannot be adequately captured through cross-country analysis without losing the institutional detail that explains the observed patterns.

The study covers the period 2010–2024; longer series (from 2000 and 2008 respectively) were used to establish productivity and firm-level comparisons. The year 2010 serves as the baseline: it coincides both with the political transition that inaugurated the policy regime examined here and with the post-financial-crisis trough of the construction sector.

The analysis draws on four categories of evidence. Macroeconomic and sectoral statistics are sourced primarily from KSH and Eurostat, including national accounts, construction output indices, employment data, the house price index, and construction producer price indices. Real output series are derived by deflating nominal output figures using sector-specific producer price indices, allowing price and volume effects to be separated.

Policy and institutional evidence is drawn from official government documents and Hungarian National Bank (MNB) reports. Secondary evidence draws on peer-reviewed academic literature as well as selected institutional and think-tank reports, the latter used primarily for contextualisation. A regional comparison with peer economies is incorporated to assess whether observed patterns are Hungary-specific or reflect broader regional trends.

Firm-level data on profitability and export performance are drawn from the CrefoPort database. For each year I selected the 1,000 largest firms by revenue (net sales), from which the construction-sector subsample was identified: between 42 and 70 firms per year, rising from around 50 in the early 2010s to 58–70 by the early 2020s, an increase that itself reflects the sector's growing weight. These firms warrant attention because of their disproportionate scale: across the period they accounted for only around 4% of construction-sector employment, yet the value added they generated ranged between roughly 12% and 35% of the sector's total gross value added, peaking in 2020–2021.

Two limitations of the sample should be acknowledged. First, the largest firms by revenue are not representative of the sector as a whole: they capture large contractors while excluding the many small subcontractors that account for the bulk of employment, so the firm-level figures reported in this study characterise the large-firm segment specifically (the segment in which the ownership changes documented in Section 6 are concentrated). Second, the composition of the subsample changes over time, so trends in average performance partly reflect which firms enter or leave the top tier rather than changes within a fixed set of firms.

The study's scope is deliberately bounded. The analysis identifies patterns across macroeconomic, firm-level, and ownership data that are consistent with the theoretical predictions. It does not claim to establish causal effects through econometric identification, nor does it isolate the precise contribution of individual policy instruments to observed outcomes.

## 3. Theoretical Framework

This section sets out the conceptual framework that structures the analysis: the key definitions, the assumed background conditions, the policy interventions under examination, the mechanisms through which they are expected to operate, and the observable patterns these mechanisms predict. The framework is presented as theory. The empirical sections that follow do not formally test it; rather, they assess whether the descriptive evidence is consistent with its predictions.

### 3.1 Key Concepts

The paper relies on four key concepts.

Extensive growth, following standard usage in growth economics, denotes expansion driven primarily by the accumulation of factor inputs rather than by improvements in the productivity with which those inputs are used. Its counterpart, intensive growth, is driven by efficiency gains (Fogel, 2009).

Efficiency is used in the sense of productive efficiency, that is, output per unit of input (Farrell, 1957; Solow, 1957). Where the text refers specifically to output per hour worked, the term labour productivity is used.

Overheating refers to a condition in which demand stimulus exceeds the short-run supply capacity of a sector, so that price increases absorb a disproportionate share of nominal demand growth relative to real output expansion (Blanchard & Galí, 2007).

Rent-seeking, established by Krueger (1974) and Tullock (1967), further discussed by Congleton et al. (2008), refers to the expenditure of resources and the use of political influence to obtain redistributive transfers rather than to create productive value. In the present context it denotes strategies by which firms secure contracts, subsidies, or market access through political connections rather than competitive performance.

### 3.2 Background Condition: Asymmetric Supply Inelasticity

The framework rests on one background condition: the short- and medium-run supply of construction output is inelastic. Three features of the sector underpin this assumption. First, construction is labour-intensive, and skilled construction labour cannot be expanded quickly; persistent labour shortages act as a binding capacity constraint. Second, the planning horizon in construction is unusually long (Kovács és mtsai., 2024), so firms can only expand capacity in response to demand they expect to persist. Third, demand stimulus that is temporary, cyclical, or politically uncertain creates weak incentives for capacity investment: a firm facing uncertain future demand does not build durable capacity to serve a boom it expects to end.

This third mechanism, however, operates unevenly. For firms without privileged access to state demand, the cyclical and politically contingent character of the boom discourages capacity investment. For politically connected firms, the expectation is different. Privileged and continuing access to public contracts lowers the perceived demand risk, making capacity expansion rational for them (a mechanism consistent with the "safety through size" dynamic documented by Laki, 2022). The evidence in Section 7 bears this out — connected firms expanded their asset base dramatically over the period while non-connected firms did not — but with a crucial qualification: that expansion was disproportionately financial and acquisitive rather than the creation of new physical capacity. Sector-level capacity therefore remained tight — sustaining the price effects the framework predicts — even as ownership concentrated. The net result is that demand stimulus was capitalised into prices and rents rather than translated into proportionate additional volume.

### 3.3 Two Interconnected Policy Interventions

The Hungarian state's post-2010 intervention in the construction market is conceptualised here as the combination of two distinct policies. Intervention A is expansionary demand policy: the large-scale channelling of public resources into construction demand, both directly through public procurement of transport, sports, and other infrastructure, and indirectly through household-side instruments such as housing purchase subsidies, subsidised loans, and preferential VAT rates.

Intervention B is preferential allocation: the selection and cultivation of politically connected domestic firms — 'national champions' — as the principal beneficiaries of state-generated demand. This connects to what Voszka (2025) terms political ownership: a category of firm, emerging at the boundary of the state and private sectors, whose creation, growth, and contraction are governed by discretionary political decisions rather than market outcomes. The broader pattern of patronage and favouritism in Hungarian public procurement is documented by Szanyi (2022), Fazekas et al. (2015), and Tóth & Hajdu (2018).

The two interventions are analytically distinct but mutually reinforcing and this feedback loop is central to the framework. Preferential allocation raises the political return on demand expansion: once a circle of favoured firms exists, additional public spending on construction becomes a mechanism for channelling resources to them, creating an incentive to expand such spending beyond what sectoral needs alone would justify. Conversely, demand expansion raises the value of being a favoured firm: the larger the politically mediated market, the greater the rents available to those with preferential access. The two policies thus form a self-reinforcing system rather than two independent measures.

### 3.4 Two Transmission Mechanisms

The suggested framework distinguishes two mechanisms through which these interventions affect sectoral outcomes, operating at different levels and leaving different empirical signatures.

The first is a market-level mechanism: subsidy capitalisation. This phenomenon was established for demand-side housing interventions under inelastic supply (Hilber & Vermeulen, 2016; Hilber, 2017; Carozzi et al., 2024) and documented in the comparable Central European setting of Croatia (Kunovac & Zilic, 2022). Demand expansion (A) under inelastic supply raises prices substantially more than real output. Its effects are general: all buyers face higher prices, and all producers enjoy higher margins, regardless of political connection. Its predicted signature is a sustained divergence between sectoral price growth and the general price level, and between nominal and real output growth.

The second is a firm-level, distributional mechanism: rent extraction. Preferential allocation (B) channels contracts to connected firms at favourable and inflated prices. Its effects are selective: rents accrue to the connected subset of firms. Its predicted signature is overpricing in public procurement, a shift in the ownership structure of the sector's leading firms towards politically connected domestic owners, and profitability concentrated in firms whose market position depends on state demand rather than competitive performance.

Distinguishing the two mechanisms matters because they are supported by different bodies of evidence and warrant different degrees of confidence. The capitalisation mechanism is directly observable in price and volume data; the rent extraction mechanism is inferred from the conjunction of procurement, ownership, and profitability patterns and is, by its nature, interpretive.

### 3.5 Predicted Consequences

The framework yields four propositions that structure the empirical sections of the paper. These are shown in Table 1.

Table 1. Theoretical proposition derived from the framework

| Concept (3.1) | Mechanism (3.4) | Proposition (3.5) | Evidence |
|---|---|---|---|
| Overheating | Subsidy capitalisation | P1: prices outpace volume | Section 4–5 |
| Rent-seeking | Rent extraction | P2: profitability decoupled from competitiveness | Sections 6–8 |
| Extensive vs. intensive growth; productive efficiency | Both, jointly | P3: productivity stagnates | Section 7 |
| — | Both, jointly | P4: uneven distribution of gains | Conclusion |

Source: own work

Proposition 1 (overheating): Sectoral prices rise persistently faster than the general price level, and nominal output growth substantially exceeds real output growth. (Examined in Sections 4-5.)

Proposition 2 (rent-seeking signature): Profitability among leading firms rises and is decoupled from competitiveness. High and rising returns coincide with weak export performance, low competition in procurement, and an ownership shift towards politically connected firms. (Examined in Sections 6-8.)

Proposition 3 (productivity stagnation). Sectoral productivity stagnates or declines despite the expansion. Where profits are available through capitalised subsidies and preferentially allocated contracts, the competitive pressure to raise efficiency is weakened; rents substitute for innovation. The sector's growth is therefore extensive. (Examined in Section 7.)

Proposition 4 (distributional outcome). The gains and losses are distributed unevenly: connected firms capture rents; the state bears inflated procurement costs and subsidy outlays; households receive subsidies whose value is partly or largely eroded by induced price increases; and the wider economy bears the long-run cost of a large, low-productivity sector absorbing labour and capital. (Addressed in the Conclusion.)

These propositions are predictions of the framework. The contribution of the empirical sections is to show that the observed trajectory of Hungarian construction after 2010 is consistent with them — and, where possible, to assess whether competing explanations account for the same patterns equally well.

### 3.6 Relation to Existing Literature

The framework connects four strands of literature that have largely developed in isolation. The housing economics literature on subsidy capitalisation establishes that demand-side interventions are absorbed into prices where supply is inelastic (Hilber & Vermeulen, 2016; Hilber, 2017; Carozzi et al., 2024; Kunovac & Zilic, 2022). It does not, however, address the political allocation of the resulting demand, treating both the stimulus and the supply constraint as exogenous.

The empirical literature on politically connected firms documents that political ties raise firm value, market share, credit access, and procurement success across a wide range of countries (Faccio, 2006; Fisman, 2001; Khwaja & Mian, 2005). It operates at the firm level and does not link favouritism to sector-level price dynamics or productivity.

A third strand, on patronage and state-business relations in Hungary and Central and Eastern Europe, documents the cultivation of a politically connected national bourgeoisie at the level of the regime and the economy as a whole (Mihályi & Szelényi, 2019; Orenstein & Bugarič, 2022; Rogers, 2024; Scheiring, 2020; Szanyi, 2022), and favouritism in Hungarian construction procurement specifically (Fazekas et al., 2015; Tóth & Hajdu, 2018). This work explains how preferential allocation operates politically, but not how it interacts with demand-side stimulus to shape sectoral prices, profitability, and efficiency.

Finally, the literature on construction productivity documents long-run stagnation across advanced economies and attributes it to measurement issues, regulatory constraints, and weak innovation (Goolsbee & Syverson, 2023; Garcia & Molloy, 2025), without considering politically supplied rents as a channel that blunts the incentive to improve efficiency.

The contribution of this paper is to integrate these strands into a single sectoral framework and to add two elements absent from each. First, the framework integrates the relationship between the two interventions: demand expansion and preferential allocation are not independent policies but a mutually reinforcing system, in which favouritism raises the political return on further demand stimulus and demand stimulus raises the value of political favour. Second, it treats supply inelasticity as partly endogenous to the policy regime itself: cyclical, time-limited, and politically uncertain stimulus discourages the long-horizon capacity investment that would otherwise relax the constraint. The result is a unified account in which overheating, concentrated profitability, and productivity stagnation emerge as joint consequences of the same intervention pair — an account that, to my knowledge, has not been articulated at the sectoral level, and for which post-2010 Hungarian construction provides a clear case.

### 3.7 Competing Explanations

The patterns predicted by the framework could in principle arise from causes unrelated to the policy mix described above. Four alternative explanations deserve consideration.

The first is post-crisis catch-up: the sector's expansion might simply reflect recovery from an unusually deep post-2008 trough, a trajectory common across Central and Eastern

Europe. The regional comparison presented in Section 4 speaks directly to this: the region shows no common trajectory. Had a shared post-crisis catch-up been the driver, one would expect convergence across these economies rather than the wide dispersion actually observed. Catch-up dynamics alone cannot explain why recovery took this form in Hungary.

The second is global cost shocks: construction input prices rose worldwide after 2020, driven by pandemic-era supply disruptions and the energy price shock, and this could account for the divergence between construction prices and the general price level. Two observations limit this explanation. The Hungarian divergence began in 2017–2019, well before the global inflationary episode, and as Table 2 shows, construction prices outpaced Hungarian inflation more persistently than global cost pressures would predict.

The third is labour shortage as an independent driver: demographic decline and emigration tightened construction labour markets across the region (Kovács et al., 2025), and shortages raise costs regardless of policy. In the present framework, however, the labour constraint is not a competing explanation but a component: it is one of the conditions that made supply inelastic (Section 3.2), and it was itself partly endogenous — the demand surge absorbed labour faster than any plausible supply response could accommodate.

The fourth is the EU funds absorption cycle. The programming periods of EU structural funds could generate investment cycles independent of domestic politics. This explanation is partially compatible with the framework — EU funds were a key financing source for Intervention A — but it cannot account for the outcomes. Poland, Czechia, Slovakia and Romania are all substantial net recipients operating under the same programming cycles, yet their construction sectors diverged sharply from one another and from Hungary's. A common external funding rhythm cannot produce opposite national outcomes. What differs across these countries is not the availability of EU money but the domestic institutions through which it is allocated — and on that dimension, corruption risk in public procurement is substantially higher in the new Member States than in Western Europe, irrespective of whether contracts are nationally or EU-financed (Tóth & Hajdu, 2025). The funds explain part of the demand expansion; they do not explain its allocation.

These alternatives are not mutually exclusive with the framework, and several operate as contributing or amplifying factors. The claim of this paper is not that they played no role, but that none of them accounts for the full configuration of outcomes: the large scale of expansion, the early and persistent price divergence, the ownership transformation, and the coincidence of rising profitability with stagnant productivity. The framework's two-intervention account is offered as the explanation most consistent with this configuration as a whole.

## 4. Macroeconomic Indicators

This section examines Proposition 1. The economic policy shift and demand-stimulating measures outlined in the introduction are clearly reflected in the macroeconomic data. I examine the sector's trajectory along four dimensions: the widening gap between production volume and prices, labour market absorption, the structural transformation of output, and the fixed capital formation.

The construction industry has become one of the fastest-growing — and most overheated — sectors of the Hungarian economy. Construction production's share of GDP rose from 5.5% in 2012 to 10.1% in 2022, while gross value added (GVA) of the construction sector increased from 4.1% in 2010 to 5.9% in 2024 (the peak year is 2022 with a figure of 6.4%). The nominal value of production more than quadrupled between 2010 and 2024. The real value of construction production grew by more than 70% until 2022, then it slightly declined in 2023 and 2024. Construction prices doubled until 2022 (compared to 2010) and then they increased by a further 60% in the next two years.

Table 2. Performance and price dynamics of the Hungarian construction sector (figures in millions of HUF)

| Year | GDP | Construction Production | Construction Production (2010 prices) | Construction Prod./GDP | Construction GVA/GDP | GDP Deflator | Construction Producer Price Change |
|---|---|---|---|---|---|---|---|
| 2010 | 27 427 573 | 1 786 611 | 1 786 611 | 6.5% | 4.1% | 2.5% | 1.1% |
| 2011 | 28 492 068 | 1 731 794 | 1 694 515 | 6.1% | 4.0% | 1.9% | 2.2% |
| 2012 | 28 919 193 | 1 587 765 | 1 524 618 | 5.5% | 3.9% | 2.9% | 1.9% |
| 2013 | 30 343 344 | 1 774 914 | 1 672 546 | 5.8% | 4.0% | 2.8% | 1.9% |
| 2014 | 32 826 770 | 2 056 152 | 1 897 712 | 6.3% | 4.2% | 3.7% | 2.1% |
| 2015 | 34 984 755 | 2 153 389 | 1 938 981 | 6.2% | 4.2% | 2.8% | 2.5% |
| 2016 | 36 311 882 | 1 813 855 | 1 588 768 | 5.0% | 3.6% | 1.3% | 2.8% |
| 2017 | 39 335 929 | 2 455 341 | 2 044 345 | 6.2% | 4.3% | 4.0% | 5.2% |
| 2018 | 43 553 984 | 3 306 919 | 2 514 500 | 7.6% | 5.1% | 4.8% | 9.5% |
| 2019 | 47 940 496 | 4 354 723 | 3 018 437 | 9.1% | 5.8% | 4.8% | 9.7% |
| 2020 | 48 807 766 | 4 339 342 | 2 800 536 | 8.9% | 5.7% | 6.4% | 7.4% |
| 2021 | 55 560 466 | 5 310 868 | 3 076 788 | 9.6% | 6.1% | 6.2% | 11.4% |
| 2022 | 65 950 424 | 6 670 651 | 3 104 066 | 10.1% | 6.4% | 14.0% | 24.5% |
| 2023 | 75 292 668 | 7 229 146 | 2 902 460 | 9.6% | 6.1% | 15.1% | 15.9% |
| 2024 | 81 447 664 | 7 676 555 | 2 910 380 | 9.4% | 5.9% | 7.6% | 5.9% |

Source: own compilation based on KSH data

The last two columns of Table 2 highlight the surge in sectoral inflation. Between 2017 and 2022 in particular, construction producer prices rose markedly faster than the general price level. In 2022, the gap was striking: 10.5 percentage points above the GDP deflator. This is the signature of the subsidy-capitalisation mechanism set out in Section 3.4. The pattern matches findings from comparable settings — Carozzi et al. (2024) on Britain's Help to Buy scheme, where prices rose but construction volumes did not in supply-constrained Greater London, and Kunovac & Zilic (2022) on Croatia. In Hungary, a 2019

MNB study estimated that two-thirds of the then-available HUF 10 million CSOK (Family Housing Allowance) subsidy had been absorbed by rising housing prices in Budapest (Banai és mtsai., 2019).

Table 3 confirms the boom from an employment perspective. Following the 2012 trough, labour intake rose continuously, peaking at 391,000 workers by 2023 — equivalent to 8.3% of national employment. In 2018–2019, labour shortages became a severe constraint on output, likely contributing further to the rise in construction prices. European Commission surveys show that 75% of firms reported labour shortages as a problem in 2019. Although this pressure eased after 2020, the indicator remained above 25% even in 2025 (Kovács és mtsai., 2025).

Table 3. Employment in Construction and the Hungarian Economy

| Year | Total employment (thousand) | Construction (thousand) | Share of construction |
|---|---|---|---|
| 2010 | 3 868 | 275 | 7.1% |
| 2011 | 3 898 | 264 | 6.8% |
| 2012 | 3 954 | 247 | 6.2% |
| 2013 | 4 025 | 248 | 6.1% |
| 2014 | 4 204 | 259 | 6.2% |
| 2015 | 4 325 | 274 | 6.3% |
| 2016 | 4 475 | 279 | 6.2% |
| 2017 | 4 546 | 304 | 6.7% |
| 2018 | 4 590 | 334 | 7.3% |
| 2019 | 4 619 | 346 | 7.5% |
| 2020 | 4 565 | 361 | 7.9% |
| 2021 | 4 610 | 375 | 8.1% |
| 2022 | 4 670 | 382 | 8.2% |
| 2023 | 4 697 | 391 | 8.3% |
| 2024 | 4 699 | 388 | 8.3% |

Source: own compilation based on KSH data

Table 4 shows how policy interventions shaped the structural composition of the sector. Before the launch of the home-building programmes in 2015, residential construction accounted for only 4–5% of total output. The CSOK effect is then clearly visible: from 2016 the share rose sharply, remaining consistently above 10% from 2017 onwards and reaching 12% that year.

Table 4. Structure of Construction Production

| Year | Production (million HUF) | Buildings total (million HUF) | Share Buildings | Share Residential | Share Non-residential | Transport Infrastructure | Pipelines & Networks | Sports & Leisure |
|---|---|---|---|---|---|---|---|---|
| 2010 | 1 786 611 | 1 174 607 | 65.7% | 5.7% | 27.4% | 15.5% | 10.1% | 0.6% |
| 2011 | 1 731 794 | 1 202 747 | 69.5% | 5.5% | 29.0% | 16.3% | 10.4% | 0.5% |
| 2012 | 1 587 765 | 1 097 521 | 69.1% | 4.8% | 28.3% | 17.4% | 11.7% | 0.5% |
| 2013 | 1 774 914 | 1 255 619 | 70.7% | 4.1% | 30.5% | 18.7% | 11.3% | 0.5% |
| 2014 | 2 056 152 | 1 484 823 | 72.2% | 4.7% | 28.3% | 23.4% | 10.8% | 0.6% |
| 2015 | 2 153 389 | 1 493 339 | 69.3% | 4.5% | 27.3% | 22.1% | 9.9% | 0.8% |
| 2016 | 1 813 855 | 1 285 374 | 70.9% | 10.5% | 32.5% | 16.9% | 7.1% | 1.2% |
| 2017 | 2 455 341 | 1 795 542 | 73.1% | 12.0% | 30.4% | 19.3% | 6.3% | 1.7% |
| 2018 | 3 306 919 | 2 348 795 | 71.0% | 10.4% | 28.3% | 21.2% | 5.6% | 2.2% |
| 2019 | 4 354 723 | 3 055 118 | 70.2% | 10.5% | 27.3% | 21.5% | 6.4% | 1.6% |
| 2020 | 4 339 342 | 3 060 724 | 70.5% | 10.8% | 29.1% | 19.6% | 6.2% | 1.5% |
| 2021 | 5 310 868 | 3 587 205 | 67.5% | 10.1% | 30.7% | 16.8% | 5.1% | 1.8% |
| 2022 | 6 670 651 | 4 409 756 | 66.1% | 11.0% | 30.7% | 14.5% | 5.2% | 1.5% |
| 2023 | 7 229 146 | 4 851 253 | 67.1% | 10.7% | 30.5% | 15.3% | 6.1% | 1.0% |
| 2024 | 7 676 555 | 5 101 167 | 66.5% | 10.7% | 30.0% | 14.3% | 5.7% | 1.2% |

Source: own compilation based on KSH data

Transport infrastructure displays cyclical behaviour. This volatility complicates capacity planning and pushes up costs, as firms must adapt to sudden swings in demand. Also notable is the retreat of network and pipeline infrastructure investment, which has placed significant pressure on energy supply systems. Taken together, the data suggest that after 2010 the Hungarian construction sector was driven by two forces operating on different timelines. Transport infrastructure expanded substantially in absolute terms and dominated the early part of the period; from 2016 onwards, housing construction — fuelled by government subsidies — became the more dynamic component, its share more than doubling. That infrastructure's share did not rise even as its absolute volume grew reflects the exceptional pace of overall sectoral expansion rather than any retreat of public investment.

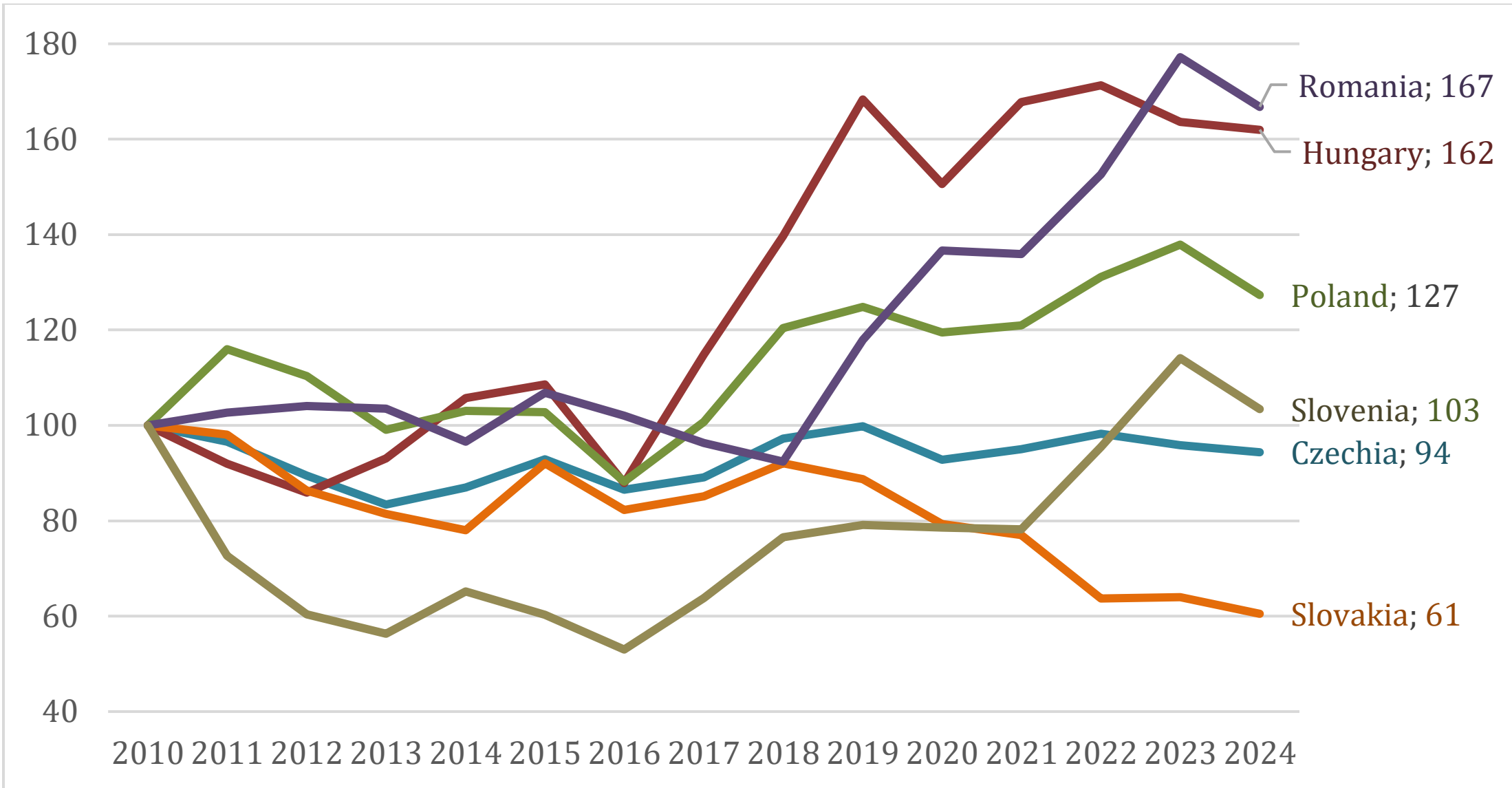

Figure 1. Volume of Construction Production in the Visegrád countries, Slovenia and Romania (real values, 2010=100)

Source: own compilation based on Eurostat data (sts_copr_a)

Figure 1 places Hungary's real construction production growth in a broader regional context, extended beyond the Visegrád group to include Slovenia and Romania. Hungary is among the region's strongest expanders: by 2024 its real construction production stood roughly 62% above its 2010 level, growth exceeded only by Romania (approximately 67%). At the opposite extreme, Slovakia is a pronounced negative outlier, its construction production contracting by nearly 40% over the same period. The region thus does not share a common trajectory: it spans a range from strong expansion (Hungary, Romania) through moderate growth (Poland) and stagnation (Czechia, Slovenia), to sharp contraction (Slovakia). The absence of a common regional pattern speaks against the hypothesis that Hungary's boom simply reflects a post-crisis catch-up trajectory shared across Central and Eastern Europe (as anticipated in Section 3.7); had catch-up been the driver, one would expect convergence rather than the wide dispersion observed.

A parallel picture emerges from investment. Gross fixed capital formation by the construction sector (Table 5), measured in national accounts terms that net out transfers of existing assets between resident firms, grew by 278% in Hungary between 2010 and 2024. That sector-wide investment expanded so strongly confirms that the capacity build-up was real rather than a statistical reflection of asset transfers between firms — a distinction that matters given the ownership consolidation documented in Section 7.

Table 5. Gross fixed capital formation by the construction sector (index, 2010=100)

| Year | Czechia | Hungary | Poland | Romania | Slovenia | Slovakia |
|---|---|---|---|---|---|---|
| 2010 | 100 | 100 | 100 | 100 | 100 | 100 |
| 2012 | 78 | 65 | 79 | 149 | 62 | 90 |
| 2014 | 77 | 118 | 111 | 81 | 64 | 102 |
| 2016 | 86 | 101 | 120 | 129 | 91 | 137 |
| 2018 | 92 | 200 | 168 | 119 | 119 | 121 |
| 2020 | 119 | 274 | 229 | 213 | 130 | 206 |
| 2022 | 228 | 374 | 183 | 352 | 204 | 148 |
| 2024 | 209 | 378 | 162 | 403 | 258 | 156 |

Source: own compilation based on Eurostat (nama_10_a64_p5), volumes rebased to 2010=100

## 5. Fiscal Intervention

In the second half of the 2010s, three groups of policy measures significantly drove demand in the Hungarian construction sector: housing subsidies, a temporary VAT reduction on new homes, and large-scale public investment in sports and transport infrastructure. Collectively, these contributed to overheating and, in certain cases, introduced cyclical fluctuations.

### 5.1 Housing Purchase Subsidies

The Family Housing Allowance (CSOK), introduced in July 2015, provided at least HUF 4,500 billion in subsidies or subsidised loans for home creation over the following decade (Makay és mtsai., 2024). The system expanded gradually and grew increasingly complex. CSOK provides non-repayable grants for families with children to buy or build new or used homes, with amounts depending on the number of children and property type. This was later complemented by the Village CSOK, launched in 2019 under the Hungarian Village Programme, which targets more than 2,600 disadvantaged settlements with fewer than 5,000 inhabitants. The scheme supports the purchase, renovation, or extension of used homes, with the explicit aim of retaining population in small settlements. At the end of 2023, the original CSOK was replaced by CSOK Plus, which provides up to HUF 50 million in subsidised loans — bridging the gap between non-repayable grants and market loans, with available amounts and repayment terms conditional on the number of children committed to having.

According to calculations by Makay et al. (2024, covering data to 2023), CSOK alone benefited 200,000 families, delivering HUF 675 billion in grants and HUF 677 billion in subsidised loans. The impact of CSOK Plus in 2024 was also substantial: MNB household loan data show that more than HUF 340 billion in subsidised housing loans were drawn that year.

The Baby-Expecting Loan (Babaváró), introduced in 2019, is an unsecured loan conditional on childbearing. It is not tied to a specific use and can be put towards home purchases, renovation, or even car purchases. Makay et al. (2024) estimate that 250,000 families used this scheme, with total support and loan volume reaching HUF 2,350 billion.

The Home Renovation Support ran in 2021–2022, with a redesigned version returning from mid-2024. The original programme offered up to HUF 3 million in non-repayable grants to families with at least one child. Its primary aim was not to expand the housing stock but to improve the quality of the existing one; it nonetheless generated substantial construction demand through renovation and modernisation works. Makay et al. (2024) estimate that around 300,000 families received a total of HUF 750 billion under the programme.

The fiscal implications of these instruments differ. Non-repayable grants burden the budget immediately, while subsidised loans carry lower upfront costs but accumulate significant fiscal risk if market interest rates remain high. What they share is their collective contribution to stimulating construction demand.

## 5.2 Reduced VAT Rate

VAT is a significant component of housing prices; at Hungary's standard rate of 27%, it substantially raises gross costs. To address this, the VAT rate on new homes was cut to a preferential 5% from 2016. The measure was initially temporary, set to expire in 2019. After a transitional period in 2020, the preferential rate returned in 2021, originally valid until 2024 but — based on current information — expected to remain in force until the end of 2026.

Temporary VAT reductions inevitably generate cyclical fluctuations. Since housing development cycles typically span 6–8 years, a short-term measure sits poorly with the timeline of multi-stage projects, making it difficult for a stable and sufficiently large supply to emerge despite the incentive (Kovács és mtsai., 2024, o. 26).

A further unintended consequence is the rise in net prices. As the 2024 expiry approached — before the extension was made public — land prices rose sharply, an increase expected to feed through into final housing prices (Kovács és mtsai., 2024, o. 26). MNB researchers reached similar conclusions about the 2016 cut: comparing prices of homes built before and after, they found that new homes became 9% cheaper overall, but the benefit was unevenly split — roughly 60% of the tax reduction was capitalised into higher net prices, with only 40% passed on to buyers (Fellner és mtsai., 2019, o. 25).

### 5.3 Public Investment

Public and EU-funded construction investment became one of the primary engines of Hungarian growth during the 2010s. The government designated construction a strategic sector, directing thousands of billions of forints towards transport and sports infrastructure. Transport investment centred on motorway expansions — including the M4, M30, M44, and M85 — and major rail projects. The most prominent of these is the Budapest–Belgrade railway modernisation, initially estimated at HUF 750 billion, with costs rising continuously since (Portfolio, 2023). Between 2016 and 2024, approximately HUF 3,200 billion in road development was completed, almost half funded from EU sources (Government of Hungary, 2023, o. 94). As Table 4 shows, transport infrastructure spending has been highly cyclical, mirroring fluctuations in housing demand — a pattern that hinders supply-side adjustment, increases uncertainty, and ultimately drives up costs.

Sport has been treated as a strategic sector in Hungary since 2010. Combined state and municipal expenditure on sports grew eleven-fold between 2010 and 2020, with total development and operational spending on facilities and training centres reaching around HUF 1,500 billion (Tóth & Mátrai, 2023, p. 89). The corporate tax support scheme introduced in 2011 channelled a further HUF 650 billion to sports organisations (Tóth & Mátrai, 2023, p. 92), much of which materialised as construction demand through infrastructure projects.

## 6. Ownership Structure

A central difficulty in any political economy account of this sector is the definition of a politically connected firm. Following Faccio (2006), a firm is treated as politically connected if a controlling owner or senior manager holds or has held public office, or maintains a documented close relationship with holders of political office. This general criterion is operationalised using two Hungarian-specific sources. The first is the classification of Tóth & Hajdu (2022), who identify a set of firms — the so-called MGTS+ group — owned or controlled by a small circle of individuals with documented ties to the governing party. The second is Laki (2022), who provides five criteria for identifying Fidesz-connected owner-managers and a named list of such individuals among Hungary's wealthiest business figures. Combining these sources, I classify the construction firms in the sample (the largest firms by revenue, drawn from the CrefoPort database) as politically connected where ownership can be traced to individuals identified in either source, supplemented by ownership records and investigative reporting for firms not covered by them. Of the firms appearing in the sample over the period, 29 meet this criterion (see the full list in the Appendix). This classification is necessarily a lower bound: political connections that operate through holding structures, nominees, or informal channels are not always visible in ownership records, so the true number of connected firms is likely higher.

This section examines Proposition 2. The ownership structure of the Hungarian construction sector has undergone a remarkable transformation (see Table 6). Whereas foreign dominance was typical before 2010, by the 2020s, Hungarian-owned firms had become decisive: by 2024 six of the ten largest firms by turnover were politically connected domestic firms, while the foreign presence took a different character than before. Of the four remaining foreign-owned firms, three were tied to specific large projects rather than to the domestic market as such. The fourth, Synergy Construction, is a Turkish-origin contractor that does not belong to the domestic connected circle but has positioned itself as a preferred builder for state-favoured foreign-investment and public infrastructure projects.

Table 6. The ten largest construction firms by turnover, 2008 and 2024

| 2008 | Ownership | 2024 | Ownership | Connected |
|---|---|---|---|---|
| STRABAG-MML | Foreign (AT) | Market Építő | Hungarian | ● |
| MASZ M6 | Foreign/Mixed | Duna Aszfalt | Hungarian | ● |
| M-Híd | Hungarian | Atomstroyexport | Foreign (RU) | |
| COLAS-HUNGÁRIA | Foreign (FR) | V-Híd | Hungarian | ● |
| Colas Hungária | Foreign (FR) | West Hungária Bau | Hungarian | ● |
| Speciális Nehézszállító | Hungarian | Bayer Construct | Hungarian | ● |
| SWIETELSKY | Foreign (AT) | Bauer Magyarország | Foreign (DE) | |
| ARCADOM | Hungarian | Samsung E&A | Foreign (KR) | |
| AKA | Foreign/Mixed | Synergy Construction | Foreign (TR) | |
| Colas Építő | Foreign (FR) | RM International | Hungarian | ● |

Source: own compilation based on CrefoPort data. ● = politically connected per the definition above. Connected firms in the top 10 by turnover: 0 in 2008, 6 in 2024.

This ownership shift is rooted in political economy. Over the past decade, the government has favoured domestic contractors in infrastructure procurement, a pattern documented in the empirical literature on Hungarian public procurement (Fazekas et al., 2015) and quantified by Tóth and Hajdu (2022). Tóth and Hajdu (2022) show that the share of total procurement value won by firms linked to the governing circle rose from around 1% to roughly 18% over the decade. The transformation is not merely a change in the nationality of owners but a reallocation of market position towards a specific, politically connected set of domestic firms. By 2024, three of the ten largest firms by turnover (V-Híd, Bayer Construct, and RM International) were linked to a single owner circle, illustrating how concentrated the sector's leading positions had become.

Table 7. Politically connected vs. other firms: profitability and competitiveness

| Year | Connected share of sector profits | ROE, connected | ROE, other | Export intensity, connected | Export intensity, other |
|---|---|---|---|---|---|
| 2008 | 15.9% | 24.2% | 15.3% | 3.3% | 6.0% |
| 2012 | 21.3% | 20.4% | 4.7% | 1.4% | 3.0% |
| 2016 | 36.9% | 35.6% | 17.4% | 0.2% | 5.2% |
| 2018 | 29.3% | 51.0% | 18.6% | 0.4% | 7.9% |
| 2020 | 31.6% | 40.3% | 15.3% | 0.3% | 0.7% |
| 2022 | 49.6% | 38.1% | 17.8% | 0.4% | 0.7% |
| 2024 | 56.1% | 41.8% | 17.3% | 0.8% | 1.3% |

Source: own compilation based on CrefoPort data. Profit share = connected firms' share of total positive pre-tax profits in the sample. ROE = after-tax profit / equity, aggregated by group. Export intensity = export revenue / turnover. Firms are classified by their post-2010 political connection status, applied across all years.

The rising prominence of connected firms is a matter of profitability, too. Table 7 tracks the connected group against the rest of the sample along three dimensions. Their share of total sector profits rose from around 16% in 2008 to 56% by 2024 — a group comprising a small minority of firms came to capture a majority of the sector's profits. This is not explained by superior efficiency in any conventional sense: connected firms' return on equity ran consistently two to three times that of other firms (41.8% versus 17.3% in 2024), a gap far larger than could plausibly be attributed to operational advantage alone, and one that widened as public demand expanded. This abnormal profitability is not an artefact of the present sample. Using ORBIS data, Oblouková et al. (2025) find Hungarian construction ROA running at roughly twice the Czech, Slovak, and Polish levels throughout 2014–2023.

At the same time, the connected firms' export intensity was consistently lower than that of other firms and fell to near zero — connected firms were oriented almost entirely towards the domestic, public-contract market rather than competitive external markets. Export orientation is, however, modest across the large-firm segment as a whole (combined export revenue never exceeded 6% of turnover even at its 2008 peak) so this indicator is best read as one signal among several. The persistent gap between the connected and other firms nonetheless remains informative of a difference in market orientation. The combination of rising profitability, exceptional returns on equity, and negligible export orientation is the signature the framework associates with rent extraction rather than competitive performance (Proposition 2).

## 7. Evolution of Efficiency

This section examines Proposition 3. A striking feature of the Hungarian construction sector is the contrast between high profitability and low productivity. In 2018, the sector's gross profit margin reached 18.5% — well above its 2010 level of 12.4% and exceeding the EU-27 average of 16.6% (European Commission, 2020).

Real labour productivity (per hour worked), however, remains below the EU average (European Commission, 2023, o. 56). Figure 2 sets Hungary's labour-productivity trajectory against its regional peers. Two observations follow. First, construction productivity stagnation is not in itself a Hungarian peculiarity. It is consistent with the literature documenting weak construction-sector productivity growth across advanced economies (Garcia & Molloy, 2023, 2025; Goolsbee & Syverson, 2023). Comparative evidence for the region points the same way: analysing technical efficiency across seven European construction sectors including Hungary and Poland, Murillo et al. (2019) find civil-engineering activities among the least efficient, with efficiency deteriorating in the years after 2010. By 2024, Hungarian productivity per hour worked had returned to approximately its 2000 level, having generated no net gain across the quarter-century. The two sharpest post-2010 productivity drops coincide with the two periods of most intensive demand stimulus: 2015–2016 (CSOK introduction and VAT reduction), and 2020–2022 (post-COVID public investment surge).

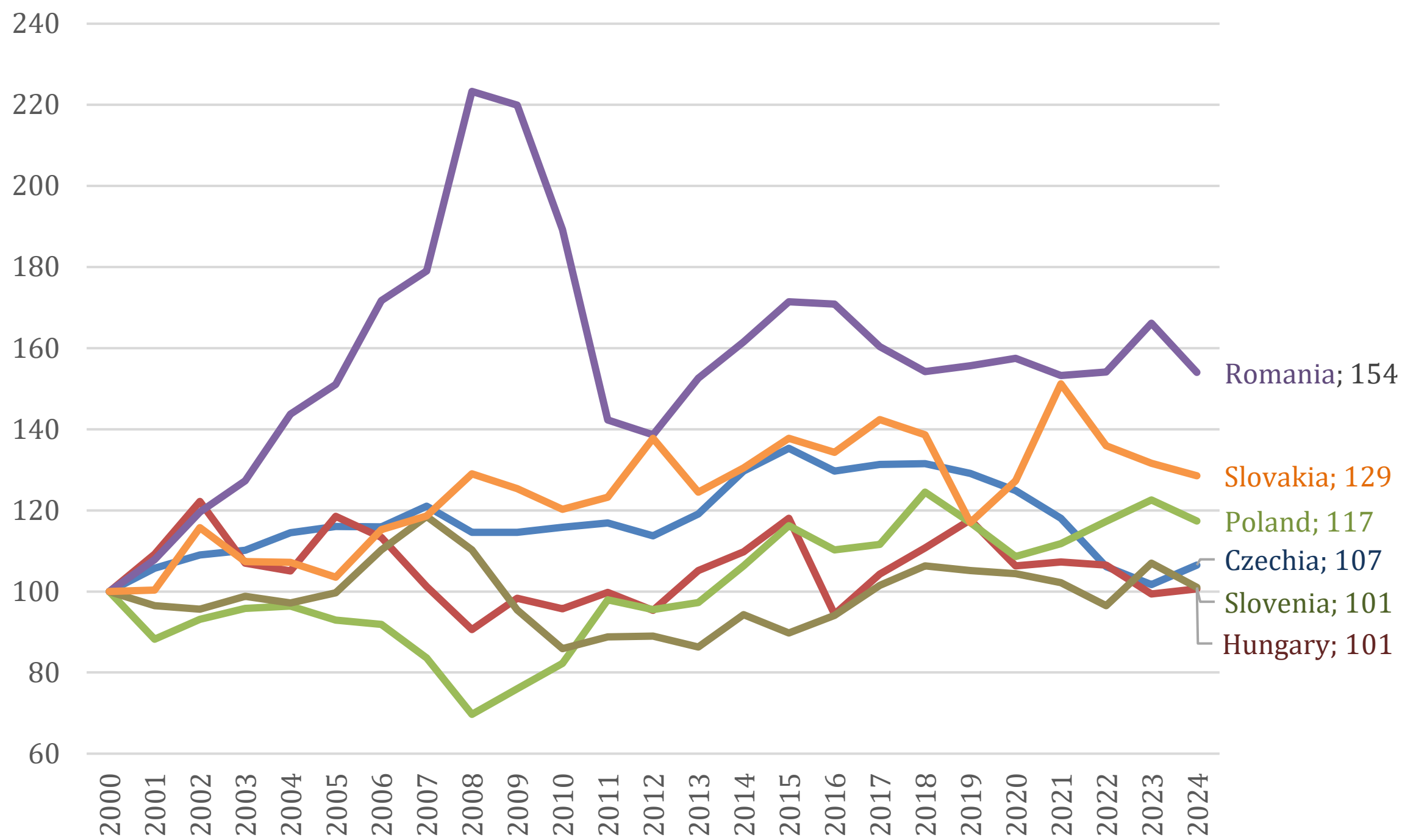

Figure 2. Real Labour Productivity Per Hour Worked (2000=100)

Source: own compilation based on Eurostat data (nama_10_lp_a21)

The second observation is the contrast between the two countries that expanded output most strongly over the period. Romania and Hungary were the region's clear outliers in both output growth (Figure 1) and investment (Table 5), yet their productivity paths diverge completely. Romanian construction productivity rose by roughly half over the period, so that its expansion was accompanied by genuine efficiency gains, the pattern one would expect of catch-up development. Hungary's was not: an exceptional expansion in output and investment coincided with flat productivity. This divergence is crucial to the paper's suggestion. It is not the scale of Hungary's construction boom that is anomalous, but the fact that so large a mobilisation of demand and capital yielded no measurable improvement in how efficiently the sector converts inputs into output.

The firm-level asset data may be used to test the inelastic supply argument presented in Section 3.2. A reasonable objection is that politically connected firms, enjoying secure access to public demand, would face little such risk and could rationally expand — and Table 8 confirms that they did. The fixed assets of connected firms grew more than twenty-fold between 2010 and 2024, while those of other firms did not grow on net at all. This last comparison should be read with caution. Because the sample comprises the largest firms by revenue in each year, its composition changes over time as firms enter and exit the top ranks; the aggregate fixed-asset figure for non-connected firms is therefore sensitive to which firms happen to be in the sample in a given year, and its year-to-year movements are correspondingly erratic. The connected group, by contrast, is a stable set of firms tracked consistently across the period, so its trajectory is the more reliable of the two. The broad contrast — sustained, order-of-magnitude growth among connected firms against no comparable trend elsewhere — is robust, even though the precise level of the non-connected series is not.

Table 8. Fixed asset growth and composition, connected vs. other firms

| Year | Fixed assets, connected (2010=100) | Fixed assets, other (2010=100) | Tangible share, connected | Tangible share, other |
|---|---|---|---|---|
| 2010 | 100 | 100 | 55% | 84% |
| 2014 | 136 | 182 | 67% | 44% |
| 2018 | 919 | 249 | 23% | 28% |
| 2020 | 1,209 | 335 | 22% | 24% |
| 2024 | 2,218 | 84 | 26% | 87% |

Source: own compilation based on CrefoPort data. Tangible share = tangible fixed assets / total fixed assets.

Taken alone, this might appear to contradict the inelastic supply premise, but the composition of the growth qualifies its interpretation. While the connected firms' tangible assets (physical plant, equipment and structures) did grow in absolute terms, their share of those firms' total fixed assets fell from 55% in 2010 to 26% by 2024. An increasing majority of connected firms' asset growth thus took the form of financial holdings and equity stakes in other firms rather than new productive capacity. At the sector level, gross fixed capital formation rose substantially (Table 5), and part of the connected firms' own growth was genuinely physical. But the drop in the tangible share indicates that connected firms expanded disproportionately through acquisition and consolidation — absorbing existing assets and market positions — relative to firms outside the connected circle.

This pattern is consistent with the framework in two ways. On the one hand, it helps explain why the sector-level supply constraint eased only slowly despite rising investment, since consolidation reshuffles existing capacity rather than adding to it. On the other hand, it reinforces the rent-seeking interpretation, since acquisition financed by contract rents entrenches market position and raises barriers to entry. This dynamic channels the returns from public demand into ownership concentration rather than into the efficiency gains that would show up as rising productivity.

## 8. Signs of Rent-Seeking

This section further examines Proposition 2. From the 2010s onwards, construction accounted for 60–80% of the value of public procurement contracts (Fazekas és mtsai., 2015) — a market widely described as a mechanism for channelling budgetary resources to politically favoured firms (Szanyi, 2022). Empirical studies confirm low competition in tenders, preferential treatment of well-connected firms (Tóth & Hajdu, 2018), and systematic overpricing (Fazekas és mtsai., 2015).

A recently published study (Bartha, 2026) examines data from Hungary's large companies to quantify the scale of rent-seeking within the economy. The research identifies three specific sectors that experienced a statistically significant and disproportionate surge in profitability between 2019 and 2023 compared to the 2008–2012 period. The construction industry stands out in this analysis: its share of total profits among large companies increased by a factor of 5.4. This finding suggests that the paradox of high profitability coupled with low productivity is driven, at least in part, by rent-seeking. In this interpretation, revenue growth derives not from superior operational performance or innovation but from politically mediated advantages in market access. Consistent with Proposition 2, the observed pattern points to rent-seeking dynamics, though the descriptive evidence cannot establish it conclusively.

Comparative evidence from the region reinforces the mechanism. Studying changes to procurement discretion thresholds in the Czech Republic, Celis Galvez et al. (2025) use bunching and regression-discontinuity designs to show that tightening procurement rules lowers final contract prices (an efficiency gain), but this gain disappears entirely for contracts awarded to politically connected firms. Political favouritism, in other words, neutralises the disciplining effect that stricter procurement rules would otherwise exert. While the Hungarian evidence in this paper is descriptive rather than causal, the Czech findings identify the mechanism the present framework posits.

The concentration of the sector points in the same direction, though it is easily understated. Conventional concentration measures applied to the largest firms are moderate and unstable over the period: the combined turnover share of the five largest firms fluctuates without clear trend, and the Herfindahl–Hirschman index remains well below levels conventionally regarded as concentrated. This apparent moderation is, however, an artefact of measuring concentration firm by firm. The politically connected firms operate under numerous separate legal identities — the appendix lists twenty-nine, of which roughly two-thirds trace to just four individuals — so measures that treat each firm as independent systematically miss the concentration that matters here. Measured by ownership rather than by legal entity, concentration rose sharply. The ownership-based profit concentration documented in Section 6 (Table 7) shows clearly what these firm-level measures fail to capture.

## 9. Conclusion

This paper examined the post-2010 expansion of Hungarian construction as a case study of large-scale government intervention, using a framework of two interconnected interventions to derive four testable propositions. The findings can be assessed proposition by proposition.

Proposition 1 (overheating) is directly supported. Real construction output grew by around 62% between 2010 and 2024, a rate matched in the region only by Romania and far above Poland, Czechia, Slovenia and Slovakia; but prices rose faster still. Construction producer prices outpaced the GDP deflator markedly from 2017 onward, peaking at a 10.5 percentage point gap in 2022, and house prices rose faster than anywhere else in the EU. That a substantial part of the subsidies was absorbed into prices rather than reaching buyers is separately estimated for both principal instruments — roughly two-thirds of the CSOK grant and 60% of the VAT reduction (Banai et al., 2019; Fellner et al., 2019).

Proposition 2 (rent-seeking signature) is supported as a pattern, though its interpretation remains inferential. Ownership of the leading firms shifted decisively from foreign to politically connected domestic hands; connected firms' return on equity ran two to three times that of other large firms throughout the period, while their exports remained negligible. Construction dominated public procurement, a domain in which low competition, preferential treatment of connected firms, and overpricing are independently documented (Fazekas et al., 2015; Tóth & Hajdu, 2018). What the data establish is the conjunction of high profitability, political connection, and weak exposure to competition; that this conjunction reflects rent extraction is an inference from the pattern, not a direct measurement of rent.

Proposition 3 (productivity stagnation) is directly supported. Labour productivity per hour worked made no net progress, returning by 2024 to roughly its 2000 level. The comparison with Romania is instructive: the region's other outlier in output and investment growth saw construction productivity rise by roughly half over the same period. Nor did the investment surge translate into productive capacity in the expected way — while sector-wide gross fixed capital formation rose by 278%, connected firms' growth was increasingly acquisitive, the tangible share of their assets falling from 55% to 26% as the balance shifted into financial holdings and equity stakes rather than new capacity.

Proposition 4 (uneven distribution) is supported in aggregate. Connected firms captured a rising share of sector profits, from around 16% to 56%; households received subsidies whose value was partly eroded by the price increases those subsidies induced; and the sector as a whole grew larger and more profitable without becoming more productive. What the evidence does not yet establish is the household-level incidence — which groups of buyers ultimately bore the capitalised subsidy.

The central interpretive claim — that these patterns jointly reflect subsidy capitalisation and politically mediated rent extraction — is a disciplined inference. The capitalisation mechanism rests on direct price and volume evidence and on well-identified findings from comparable settings (Fellner et al., 2019; Kunovac & Zilic, 2022; Carozzi et al., 2024). The rent-extraction mechanism rests on the conjunction of procurement, ownership, and profitability patterns that matches the international signature of politically connected firms (Fisman, 2001; Faccio, 2006) and the independent estimate that construction profits grew disproportionately among Hungarian large firms (Bartha, 2026). The competing explanations considered in Section 3.7 — post-crisis catch-up, global cost shocks, labour shortages, and the EU funds cycle — each could account for part of the picture, and several amplified the dynamics described here. None of them, however, explains why recovery took this exceptional form only in Hungary, why prices diverged before the global inflationary episode, or why the gains concentrated in politically connected firms.

The analysis has clear limitations. It is descriptive: it establishes consistency with the framework rather than identifying causal effects, and it cannot quantify how much of the price divergence is attributable to each policy instrument, nor how much of the profitability surge constitutes rent rather than return. The firm-level evidence covers only the largest companies, and its composition changes over time. Political connection is identified from ownership records and the published classifications of connected individuals; while this yields a documented and reproducible list, it captures only connections visible in ownership data and so represents a lower bound.

These limitations define the open questions. A counterfactual exercise — what housing output the subsidy outlays would have purchased under elastic supply — could quantify the welfare cost of capitalisation. Measuring the household-level incidence of the eroded subsidies — which groups of buyers ultimately bore the capitalised cost — would require linked transaction and subsidy data, and deserves a study of its own.

The analysis carries direct implications for policy design. First, demand-side housing and infrastructure programmes should be conditioned on the supply capacity of the targeted sector: where supply is inelastic, subsidies are predictably capitalised into prices, and the larger the stimulus, the greater the share absorbed by sellers rather than the intended beneficiaries. Stimulus paired with supply-side measures — workforce development, predictable long-term programming, stable regulatory horizons — is less likely to dissipate into prices. Second, the temporal design of interventions matters as much as their size: temporary and cyclical instruments, such as time-limited VAT preferences and electorally timed investment surges, discourage the capacity investment that would make the stimulus effective, so credible multi-year commitments should be preferred to renewable short-term schemes. Third, the effectiveness of any demand expansion depends on the integrity of its allocation: where procurement competition is weak, additional public spending raises the rents available to connected firms rather than output, implying that competitive tendering safeguards and transparency requirements are not merely governance concerns but preconditions for the stimulus to reach its target. Finally, for EU policy specifically, the findings suggest that the volume of cohesion funding matters less than the procurement institutions through which it flows — strengthening conditionality on competitive allocation may do more for real outcomes than adjusting funding levels.

The broader lesson extends beyond Hungary. Demand-side stimulus and politically structured procurement are common policy instruments; what the Hungarian construction sector illustrates is their interaction. Demand expansion under inelastic supply inflates prices; preferential allocation converts inflated prices into concentrated rents; and rents, once available, weaken the competitive pressure that drives productivity — leaving a sector that is larger, more profitable, and no more efficient than before. Where these instruments are combined, extensive growth without efficiency is not an anomaly but a predictable outcome. The framework also implies a forward-looking test: because it ties the sector's trajectory to a specific configuration of demand policy and preferential allocation, a change in that policy regime should, over time, register in the sector's prices, ownership concentration, and procurement patterns. This makes the coming years a natural test of whether the mechanisms identified here are as central as the framework proposes.

Data sources

Crefoport commercial database of Hungarian company financial reports: https://www.crefoport.hu/scholar

Eurostat construction sector production data: https://doi.org/10.2908/STS_COPR_A

Eurostat gross fixed capital formation by industry: https://doi.org/10.2908/NAMA_10_A64_P5

Eurostat housing price statistics: https://ec.europa.eu/eurostat/web/housing-price-statistics/database

Eurostat labour productivity data: https://doi.org/10.2908/NAMA_10_LP_A21

KSH (Hungarian Central Statistics Office) construction sector data: https://www.ksh.hu/stadat?lang=hu&theme=epi

KSH (Hungarian Central Statistics Office) national accounts data: https://www.ksh.hu/stadat?lang=hu&theme=gdp

MNB (Hungarian National Bank) household loan statistics: https://statisztika.mnb.hu/sk-haztartasi

Appendix. Firms classified as politically connected

The 29 firms below are those treated as politically connected in the calculation. Classification follows the operational definition in Section 6: a firm is included where ownership can be traced to individuals identified by Tóth & Hajdu (2022) or Laki (2022), or, for firms not covered by those sources, where ownership records and investigative reporting document a connection. One firm (Metal Hungária Holding) is included as a probable case, connected on weaker or indirect evidence. Distinct legal entities under common ownership are listed separately, as they appear separately in the firm-level data.

| Firm | Controlling owner / circle | Basis of classification |
|---|---|---|
| Mészáros és Mészáros (Kft.) | Lőrinc Mészáros | Tóth & Hajdu (2022); Laki (2022) |
| Mészáros és Mészáros (Zrt.) | Lőrinc Mészáros | Tóth & Hajdu (2022); Laki (2022) |
| Mészáros M1 Útépítő | Lőrinc Mészáros | Ownership records |
| R-Kord Építőipari | Lőrinc Mészáros | Tóth & Hajdu (2022) |
| Fejér-B.Á.L. Építő | Lőrinc Mészáros (family) | Tóth & Hajdu (2022) |
| ZÁÉV Építőipari | Lőrinc Mészáros (since 2019) | Tóth & Hajdu (2022) |
| RM International | Lőrinc Mészáros | Ownership records (HQ Felcsút) |
| V-Híd Építő | Lőrinc Mészáros & Zsolt Homlok | Tóth & Hajdu (2022) |
| V-Híd Network | Lőrinc Mészáros & Zsolt Homlok | Tóth & Hajdu (2022) |
| Vasútvillamosító | Zsolt Homlok (Mészáros's son-in-law) | Tóth & Hajdu (2022) |
| Duna Aszfalt (Kft.) | László Szíjj | Tóth & Hajdu (2022) |
| Duna Aszfalt (Zrt.) | László Szíjj | Tóth & Hajdu (2022) |
| Duna Aszfalt Projekt | László Szíjj | Tóth & Hajdu (2022) |
| Duna Group Európa Útépítő | László Szíjj | Ownership records (same group) |
| Hódút Építő | Károly Varga & László Szíjj | Tóth & Hajdu (2022) |
| Vakond Via | Károly Varga & László Szíjj | Tóth & Hajdu (2022) |
| Magyar Vakond | Károly Varga & László Szíjj | Tóth & Hajdu (2022) |
| Market Építő | István Garancsi | Tóth & Hajdu (2022) |
| West Hungária Bau | Attila Paár | Tóth & Hajdu (2022) |
| Magyar Építő | Attila Paár (since 2016) | Tóth & Hajdu (2022) |
| Nemzetközi Vegyépszer (entity 1) | L. Simicska (to 2018); later Elek Nagy | Tóth & Hajdu (2022); Laki (2022) |
| Nemzetközi Vegyépszer (entity 2) | L. Simicska (to 2018); later Elek Nagy | Tóth & Hajdu (2022); Laki (2022) |
| Bayer Construct | Attila Balázs (Tiborcz/Mészáros ties) | Ownership records; media |
| Vabeko | Bernula family (Mészáros partnership) | Ownership records; media |
| VER-BAU | Sz. Benkő & R. Weininger (Szijjártó ties) | Ownership records; media |
| Soltút Útépítő | Kálmán Rencsár | Ownership records; media |
| Szabadics Építőipari | Szabadics family | Ownership records; media |
| Raw Development | Bálint Somlai (Matolcsy circle) | Ownership records; media |
| Metal Hungária Holding (probable) | József Kreinbacher | Media (indirect) |